\documentstyle[eqsecnum,aps]{revtex}

\begin{document}
\draft
\preprint{HEP/123-qed}
\title{Mass Enhancement in Narrow Band Systems}
\author{Marco Zoli}
\address{Istituto Nazionale di Fisica della Materia - 
Dipartimento di Matematica e Fisica, \\  Universit\'a di Camerino, 
62032 Camerino, Italy. e-mail: zoli@campus.unicam.it
}

\date{\today}
\maketitle
\begin{abstract}
A perturbative study of the Holstein Molecular
Crystal Model which accounts for lattice structure
and dimensionality effects is presented. Antiadiabatic
conditions peculiar of narrow band materials and
an intermediate to strong electron-phonon coupling
are assumed. The polaron effective mass 
depends crucially in all dimensions on the intermolecular 
coupling strengths which also affect the size of the
lattice deformation associated with the small
polaron formation. 

\end{abstract}
\narrowtext
\section*{I.Introduction}
Several studies published over the last years have
addressed the questions related to the existence
of small polarons with itinerant properties in real
systems \cite{eminholst,eminran,kostur,io,aubry,haki,chris}. This issue is central for a possible description
of high $T_c$ superconductivity in terms of
(bi)polaronic models \cite{alexran,krebs92}. In spite of being well defined
quasiparticles, small polarons may loose their mobility
either because of a dynamical dephasing between the charge
carriers and their surrounding deformation field or because of the
heaviness of the effective mass. These effects could
however differ significantly according to the regime
(adiabatic or antiadiabatic) and the strength of electron-
phonon coupling characterizing the system \cite{alenew}. 
Theoretical investigations start generally from the
Holstein Molecular Crystal Model \cite{holstein59},
a fundamental tool which has revealed a rich variety
of behaviors in the polaron landscape through the
use of quantum Monte Carlo \cite{raedt,kornilo}, density matrix renormalization group techniques 
\cite{jeckelmann}, variational
methods \cite{chris,toyozawa,pucci,tsiro}, cluster solutions \cite{ranthib,mustre,alekab,fehske} and perturbative approaches \cite{gogolin,stephan,grilli}.
A numerical study of the polaron bandwidth in the
first order of perturbative theory has proved that
the phonon momentum dependence is a {\it key} feature
of the Holstein Hamiltonian and that the lattice
dimensionality strongly influences the bandwidth values
\cite{prb98}.
Unlike other properties such as ground state energy
and effective mass, the bandwidth is not affected
by second order corrections and therefore
it provides a useful testing bench for alternative, non
perturbative attacks on the polaron problem \cite{robin}.
Being aware of the importance that the intermolecular
forces have in the narrowing of the polaron band,
we report here on a perturbative numerical study of the
mass enhancement in the strong coupling and antiadiabatic regime. The reasons why I choose to start from this
regime are threefold: i) it is the easiest in the sense that the lattice deformation follows
coherently the charge carriers and the abovementioned
dephasing features can be ruled out, ii) the unit
comprising electron {\it and} phonon dressing is a
stable small polaron that is, the size of the quasiparticle
is not significantly broadened
in some portions of our parameter space, iii) this
regime is relevant to several classes of narrow band
materials whose charge
carriers effective masses deserve accurate estimates.
Although the present work assumes that the  carriers
are coupled to bosonic degrees of freedom having a
vibrational origin, antiadiabatic conditions are
likely to occur in excitonic systems where the
characteristic frequency $\bar \omega$ could be easily
of order of $1eV$
and the boson can therefore follow the electron
eesentially without retardation. In these cases the
carriers effective mass is expected to be only moderately enhanced with
respect to the bare electron mass.

In Section II, the dispersive Holstein model is
briefly reviewed while the numerical results are
displayed in Section III. Some conclusions are
drawn in Section IV.

\section*{II. Holstein Model with Dispersive Phonons}

My starting point is the real space - momentum space
representation of the Holstein Hamiltonian which 
reads
\begin{eqnarray}
H=& & -t \sum_{i \ne j }c_i^{\dag}c_{j}
+ \epsilon \sum_j c_j^{\dag}c_{j}
+ \sum_{\bf k} \hbar \omega_{\bf k}a^{\dag}_{\bf k}
a_{\bf k} \,
\nonumber \\
& &+ {g \over \sqrt{N}} \sum_{\bf k} \sum_j c_j^{\dag}c_j (a_{\bf k} + a_{-{\bf k}}^{\dag})exp \bigl(i{\bf k} \cdot {\bf r}_j \bigr)
\label{1}
\end{eqnarray}
 $c_i^{\dag}$ ( $c_{i}$ ) creates
(destroys) a tight binding electron at the $i-th$ molecular
site
and $t$ is the hopping integral related to the
bare electron half bandwidth $D$ by $D=zt$, $z$ being
the coordination number. $\epsilon$ is a reference
electronic energy and ${\bf r}_j$ is the $j-th$ lattice site vector. $N$ is the number of molecules in the lattice. It is understood that
$t$ differs from zero only between  first neighbors sites
and this poses a constraint on the $i \ne j$ sum
in the first addendum.
$a_{\bf k}^{\dag}$ ( $a_{\bf k}$ ) creates
(destroys) a {\bf k}-phonon with frequency $\omega_{\bf k}$. The lattice dimensionality enters the problem through
the phonon dispersion relations which have been
obtained  analytically by assuming first neighbors
pairwise intermolecular
forces both along the linear chain (1D), the square lattice
(2D) and the simple cubic lattice (3D) \cite{ijmp}:

\begin{eqnarray}
& &\omega^2_{1D}(k)= {{\beta + \gamma } \over M} 
+ {1 \over M} \sqrt { \beta^2 +  A_{x} } \,
\nonumber \\
& &\omega^2_{2D}({\bf k})= 
{{\beta + 2 \gamma } \over M} + 
{1 \over M} \sqrt { \beta^2 + 2B_{x,y}} \,
\nonumber \\
& &\omega^2_{3D}({\bf k})= 
{{\beta + 3 \gamma } \over M} + 
{1 \over M} \sqrt { \beta^2 +  C_{x,y,z}} \,
\nonumber \\
& &A_{x}= \gamma^2 + 2 \gamma \beta  c_x
\,
\nonumber \\
& &B_{x,y}=  \gamma ^2 (1 + c_xc_y + s_xs_y) + 
\beta \gamma ( c_x + c_y ) 
\,
\nonumber \\
& &C_{x,y,z}= \gamma ^2 (3 + 2(c_xc_y + s_xs_y + 
c_xc_z + s_xs_z   
\,
\nonumber \\
& & + c_yc_z + s_ys_z) ) + 2 \beta 
\gamma (c_x + c_y + c_z) 
\label{2}
\end{eqnarray}

where, $c_x=cosk_x$, $c_y=cosk_y$, $c_z=cosk_z$,
$s_x=sink_x$ etc.
$\beta$ is the intra-molecular force constant and $\gamma$ 
is the inter-molecular first neighbors 
force constant. 
 Let's define $\omega_0^2=\,2\beta/M$ and $\omega_1^2=\,\gamma/M$
with $M$ being the reduced molecular mass.
$N$ is the number of diatomic molecules in the lattice
and $g$ is the local electron-phonon coupling 
constant.
The adiabatic parameter is $\hbar \bar \omega/D$, 
$\bar \omega$ being a characteristic phonon frequency
which we take as the zone 
center frequency and whose expression is: 
$\bar \omega^2=\, \omega_0^2 + z\omega_1^2 $.

Throughout this paper
we fix $\hbar \omega_0=\,100meV$ and $t=15meV$ so that
the antiadiabatic condition $\hbar \bar \omega/D > 1$
is fulfilled in any
dimensionality. Moreover, our perturbative approach
 requires the occurence of the condition
$D < g$ \cite{eagles}.
By applying the Lang-Firsov unitary transformation
\cite{lang} and the subsequent $1/\lambda_0$
expansion with $\lambda_0 \equiv g^2/(\hbar \omega_0 D)$
being the ratio between polaron binding energy and 
electron half bandwidth \cite{firsov}, 
$H$ of eq.(1) transforms into $\tilde H=\,
{\tilde H}_0 + {\tilde H}_P$ with: 
\begin{eqnarray}
{\tilde H}_0& &= \sum_{\bf k} \hbar \omega_{\bf k}a^{\dag}_{\bf k}a_{\bf k}
+ \bigl(\epsilon - {{g^2} \over {N}}\sum_{\bf k} (\hbar \omega_{\bf k})^{-1} \bigr) \sum_j c_j^{\dag}c_{j} \,
\nonumber \\
& & - {{g^2} \over {N}}\sum_{\bf k} \sum_{i \ne j }
{{ exp \bigl(i{\bf k} \cdot ({\bf r}_i - {\bf r}_j) \bigr)}  \over {\hbar \omega_{\bf k}}} c_j^{\dag}c_{j}c_i^{\dag}c_{i} \,
\nonumber \\
{\tilde H}_P& &= -t \sum_{i \ne j }exp\Bigl[ 
- {{2g^2} \over {N}}\sum_{\bf k}(\hbar \omega_{\bf k})^{-2}
sin^2 { {({\bf k} \cdot ({\bf r}_i - {\bf r}_j))} 
\over 2} \Bigr] \cdot \,
\nonumber \\
& &\sum_{m=0}^{\infty} {1 \over {m!}}
{ \Bigl[
{g \over \sqrt{N}}\sum_{\bf k} {{ a^{\dag}_{-{\bf k}}} \over {\hbar \omega_{\bf k}} } \bigl(
e^{ i{\bf k} \cdot {\bf r}_i } -
e^{ i{\bf k} \cdot {\bf r}_j } \bigr) \Bigr]^m } 
\cdot \,
\nonumber \\
& &\sum_{n=0}^{\infty} {1 \over {n!}}
{ \Bigl[ 
{g \over \sqrt{N}}\sum_{\bf k} {{ a_{{\bf k}}} \over {\hbar \omega_{\bf k}} } \bigl(
e^{ i{\bf k} \cdot {\bf r}_j } -
e^{ i{\bf k} \cdot {\bf r}_i } \bigr) \Bigr]^n } 
c_i^{\dag}c_{j}
\label{2}
\end{eqnarray}
${\tilde H}_0$ is diagonal except for a second
order term in the electron density operator which could
cause an attractive
electron-electron interaction \cite{alexran}. The perturbation ${\tilde H}_P$ displays the fundamental features of the
polaronic quasiparticle as the hopping integral narrowing (first factor in eq.(3)) plus the peculiar mixing of fermionic and
bosonic variables. At any electron-phonon interaction
vertex $m$ ($n$) phonons can be emitted from (absorbed by)
the cloud surrounding the electron provided the total
crystal momentum is conserved.
By choosing a transformed
ground state with no phonons we see that the first
order dispersive contribution $E^{(1)}$ to the ground 
state polaron band
arises only from the $n=m=0$ term in eq.(3) hence from
the zero phonon scattering process.
In 3D and taking a lattice spacing 
$a=\,|{\bf r}_i - {\bf r}_j | =1$, one finds
\begin{eqnarray}
E^{(1)}({\bf p})& &=\, 
-2t(\cos p_x + \cos p_y + \cos p_z) \cdot
\,
\nonumber \\
& &exp\Bigl[ 
- {{2g^2} \over {N}}\sum_{k_x} \sin^2 {{k_x } \over 2}
\sum_{k_x, k_y} (\hbar \omega_{\bf k})^{-2}
 \Bigr]
\label{3}
\end{eqnarray}
where the total crystal momentum  {\bf p} coincides
with the electron momentum due to the absence of
self-energy corrections.
The second order perturbative contribution requires
summation over all intermediate states having
$m$ {\bf k}-phonons more than the vacuum and one electron
on a $i$ first neighbor of the $j$ initial site. Moreover, the final electronic position $f$ can either coincide with $j$
(this process does not introduce dispersive effects
in the polaron band) or be a first neighbor of the $i$
site. The latter event is clearly dimension dependent:
in 1D the final site is a second neighbor of $j$, 
in 2D $f$ can be either a second or a third neighbor
of $j$ and in 3D, also the fourth neighbor site can
be reached via hopping. While
the detailed study of these dispersive effects (which
can become relevant in adiabatic conditions) is  
postponed to a next paper we turn now to compute
the polaron effective mass. It should be remarked
that the second order corrections decrease the mass
values calculated in first order perturbative theory
which therefore should be meant to provide upper
bounds for the polaron mass.

\section*{III. Polaron Effective Masses}

The polaron mass $m^*$ can be obtained
according to the
definition
\begin{equation}
{{m^*} \over {m_0}}=\, 
{{zt} \over {\nabla^2 E({\bf p})|_{{\bf p}=0}}}
\label{4}
\end{equation}
where $m_0$ is the bare band mass and the dispersive
polaron band is given by eq.(4). The polaron binding 
energy has obviously no {\bf p}-dispersion. Then,
$m^*/m_0$ is at first order independent of $t$ and
coincident simply with the reciprocal of the band
narrowing factor. This picture holds in the strong
coupling regime here assumed.
In Fig.1(a), the ratios $m^*/m_0$ for the 1D, 2D and 
3D are computed versus the first neighbor intermolecular
force constant $\omega_1$.
While the polaron masses strongly depend on the dimensionality $d$ and are very large at small $\omega_1$, the ratios
become essentially $d$ independent in the upper portion
of the parameter range and tend to converge to 2.
The value of the polaron binding energy 
$\lambda_0  > 1$ signals that 
the energy gain associated with the lattice deformation
is larger than the kinetic energy due to the tight
binding hopping in the bare band. Therefore it is energetically convenient to the electron to be dressed
by the phonon cloud and become a
quasiparticle. Actually, in antiadiabatic regimes,
the more restrictive condition concerning the lattice
deformation
$\alpha_0 \equiv g/(\hbar \omega_0) > 1$ needs to
be fulfilled to guarantee that our quasiparticle
{\it is a small polaron}. While $\lambda_0$ and
$\alpha_0$ refer to a system with dispersionless
phonons it is clear that both polaron binding energy
and lattice deformation parameter will change after
switching on the intermolecular interactions.
The role of the intermolecular couplings is not simply
that of increasing the characteristic phonon frequency but rather that of establishing the correct Holstein model dependence on dimensionality. Ignoring the intermolecular
couplings would yield the 1D polaron band $\Delta E_{1D}$ larger than
the 2D polaron band $\Delta E_{2D}$ and 
$\Delta E_{2D} > \Delta E_{3D}$ which is clearly
unphysical since the polaronic wave functions overlap
is larger in higher dimensionality. This wrong trend
would hold for any value of the intramolecular frequency
$\omega_0$. Then, as observed by Holstein himself in 
his original paper \cite{holstein59}, the phonon dispersion
is a vital ingredient of the theory and this observation
motivates our numerical investigation.

Because of our definitions $\lambda_0$ does not 
depend on $d$
whereas $\alpha_0$ is $\propto \sqrt{d}$.
In Fig.1(b) we see that the lattice deformation 
$\alpha=\, N^{-1} g \sum_{\bf k} 
{\hbar \omega_{\bf k}}^{-1}$ is
in all dimensions a decreasing function of the
intermolecular force constant and, in 1D, the system
does not fulfill the small polaron condition at
the largest $\omega_1$ values. This case has been 
presented to point out how the starting condition
$\alpha_0^{1D}=1.3$ sets the 1D system rather
in an intermediate coupling regime where 
a broadening
of the polaron size can take place \cite{kopidakis,kabanov}. Under these conditions
the same perturbative method based on the Lang Firsov transformation becomes questionable \cite{io1}.
Below $\bar {\omega}_1=\,48meV$ the polaron bandwidth
inequalities $\Delta E_{3D} > \Delta E_{2D} > \Delta E_{1D}$ 
are not satisfied \cite{prb98} as expected on general grounds 
hence, the dispersionless and
the weakly dispersive Holstein Hamiltonian 
yield erroneous estimates of the effective masses.
The straight line in Fig.1(b) marks therefore the lower
bound for the intermolecular coupling which guarantees
the validity of the model.

\begin{figure}
\vspace*{14truecm}
\caption{(a) One, two and three dimensional polaron 
masses (in units of the bare band mass) 
versus the first neighbors
intermolecular energy. The dispersionless polaron binding energy is 5.3, in units of the bare
electron kinetic energy $D=\,zt$.
(b) One, two and three dimensional 
lattice deformations 
versus the first neighbors
intermolecular energy. The $\alpha_0^d$ are the lattice deformation values in a dispersionless model ($\omega_1=\,0$).
$\bar {\omega}_1=\, 48meV$ marks the lower 
bound for the validity of the model
(see text).
\label{autonum}}
\end{figure}

\begin{figure}
\vspace*{14truecm}
\caption{(a) As in Fig.1(a) but with a dispersionless polaron binding energy $\lambda_0=\,10.9$. (b) As in Fig.1(b) but with larger lattice deformation values.
The lower 
bound for the validity of the model is set at $\bar {\omega}_1=\, 59meV$.
\label{autonum}}
\end{figure}

Increasing the electron-phonon coupling, see Fig.2(a),
leads to a strong mass enhancement (particularly in 3D)   at small $\omega_1$ while the mass ratios converge
to 5 at large intermolecular coupling strenghts.
Note (Fig.2(b)) that the polaron is small in all $d$ throughout the whole $\omega_1$ range hence the Lang Firsov
method works well in this case. The
threshold value for the validity of the Holstein model
is set here at $\bar {\omega}_1=\,59meV$.
The inequalities $m^*_{3D} > m^*_{2D} > m^*_{1D}$ keep on being satisfied for a portion
of $\omega_1$ values above the threshold before
convergence is achieved but second order perturbative
terms
(being larger in higher dimensionality) are expected to correct partly this trend.
Figs.3 show that a stronger {\it e-ph} coupling,
with $\lambda_0=21.3$, yields a mass ratio of $\simeq 25$
and shifts the threshold $\bar {\omega}_1$ at $65meV$ 
pointing out the relationship between features of the
phonon spectrum
and strength of the $g$ coupling. Also in this case
the polaron size remains {\it small} throughout the
whole $\omega_1$ range (see Fig.3(b))
thus confirming the reliability
of the Lang Firsov method in a strong {\it e-ph}
coupling regime with antiadiabatic conditions.

\begin{figure}
\vspace*{14truecm}
\caption{(a) As in Fig.1(a) but with a dispersionless polaron binding energy $\lambda_0=\,21.3$. (b) As in Fig.1(b) but with larger lattice deformation values.
The lower 
bound for the validity of the model is set at $\bar {\omega}_1=\, 65meV$.
\label{autonum}}
\end{figure}

Next, I have varied
$g$ in the range 1 - 4 (in units of $\hbar \omega_0$
and found, at any $g$,  the minimum intermolecular coupling $\bar {\omega}_1 (g)$ at which the
bandwidth inequalities $\Delta E_{3D} > \Delta E_{2D} > \Delta E_{1D}$ are satisfied. This criterion yields 
an empirical relation, $\bar {\omega}_1 (g) \simeq
\bar {\omega}_1 (1)(1 + ln(g))$, which allows one to  obtain a reliable estimate of the polaron effective mass.  In Fig.4 the 1D mass ratio is plotted versus the
dimensionless $g/(\hbar \omega_0)$ both in the
first and second order of perturbative theory:
it turns out that second order corrections are
negligible in 1D systems with antiadiabatic
conditions as those we have assumed. 
I want to point out that the mass values reported 
in Fig.4 correspond, at any $g$,
to the minimum $\omega_1$ (the threshold) which ensure
the smallness of the ground state polaron. Then, they are
upper bounds for the 1D mass in the sense that the
presence of larger intermolecular forces would yield
lighter mass values.
As expected on general grounds \cite{lowen,gerlow}
the small polaron
solution is the ground state of the discrete
Holstein model in the intermediate to strong
{\it e-ph}-coupling regime here considered while, by decreasing
the coupling, a continuous
cross over to large polaron solutions can take place
in 1D \cite{kopidakis}. We have however seen (Fig.1(b))
that the dispersive features of the phonon spectrum could affect
the transition by inducing a spreading of the lattice
deformation. 
Anyway, the smoothness of our $m^*$ versus $\alpha_0$ curve
(persisting also in the lower $\alpha_0$ range not displayed in Fig.4) confirms that no self trapping 
is found
in 1D antiadiabatic regimes whereas recent 
variational \cite{romero} and perturbative \cite{io1}
investigations  signalling a rapid growth of $m^*$ vs.  {\it e-ph}
coupling  support the
existence of the self trapping transition between polaron
states of different structure in 1D adiabatic
systems. In any case, phase transitions in Holstein models are ruled out being the ground state energy
an analytic function of the {\it e-ph} coupling.

\begin{figure}
\vspace*{7truecm}
\caption{ One dimensional polaron 
mass (in units of the bare band mass) 
versus the lattice deformation parameter.
Both the first and second order perturbative
results are displayed.
\label{autonum}}
\end{figure}

\section*{IV. Concluding Remarks}

I have presented the first results of a perturbative 
approach to the polaron problem which focusses on
the lattice dimensionality effects. Having chosen
the antiadiabatic regime of the Holstein Molecular
Crystal Model we are confident of the accuracy 
of the first order perturbative theory for
one dimensional systems with strong {\it e-ph}
coupling whereas some significant second order 
corrections can
occur in higher dimensionality \cite{io1}. 
While a previous
work \cite{prb98} had shown that a dispersionless Holstein model
leads to i) erroneous estimates of the polaron bandwidth
versus dimensionality and ii) unphysical divergences 
in the site jump probability \cite{yama}, the present study reveals
that the polaron effective mass 
is in all dimensions
very sensitive to the strength of the 
forces which tie the molecules in the lattice.
We obtain polaron masses between 2 and 25 times the bare band mass by varying the {\it e-ph} coupling
in the range $(\simeq 1 - 2.5)$ and these values become
essentially dimension independent when the intermolecular forces
are sufficiently strong. 
The 
molecular lattice structure has been described
by means of a single
parameter, the first neighbor intermolecular coupling,
being understood that the range of the interactions
should be extended in real systems if least squares
fitting of the experimental phonon frequencies can 
provide effective values for the next neighbors and
long range force constants \cite{zoli}.
The antiadiabatic regime
with strong {\it e-ph} coupling
ensures the validity of the quasiparticle picture
for the small polaron nonetheless we have seen that some
broadening  of the phonon cloud can arise at intermediate {\it e-ph} couplings for strong values of the intermolecular forces with consequent
lowering of the lattice deformation parameter.
This interesting feature suggests that the intermolecular
forces influence the quasiparticle size and, 
incorporating the effects of the
{\it e-ph} coupling, have a role in driving the continuous
transition between large and small polaron.


\begin{references}
\bibitem{io1}
The questions related to the validity of the perturbative approach both in the adiabatic and antiadiabatic
regime will be discussed in detail in a forthcoming
paper.
\end{references}
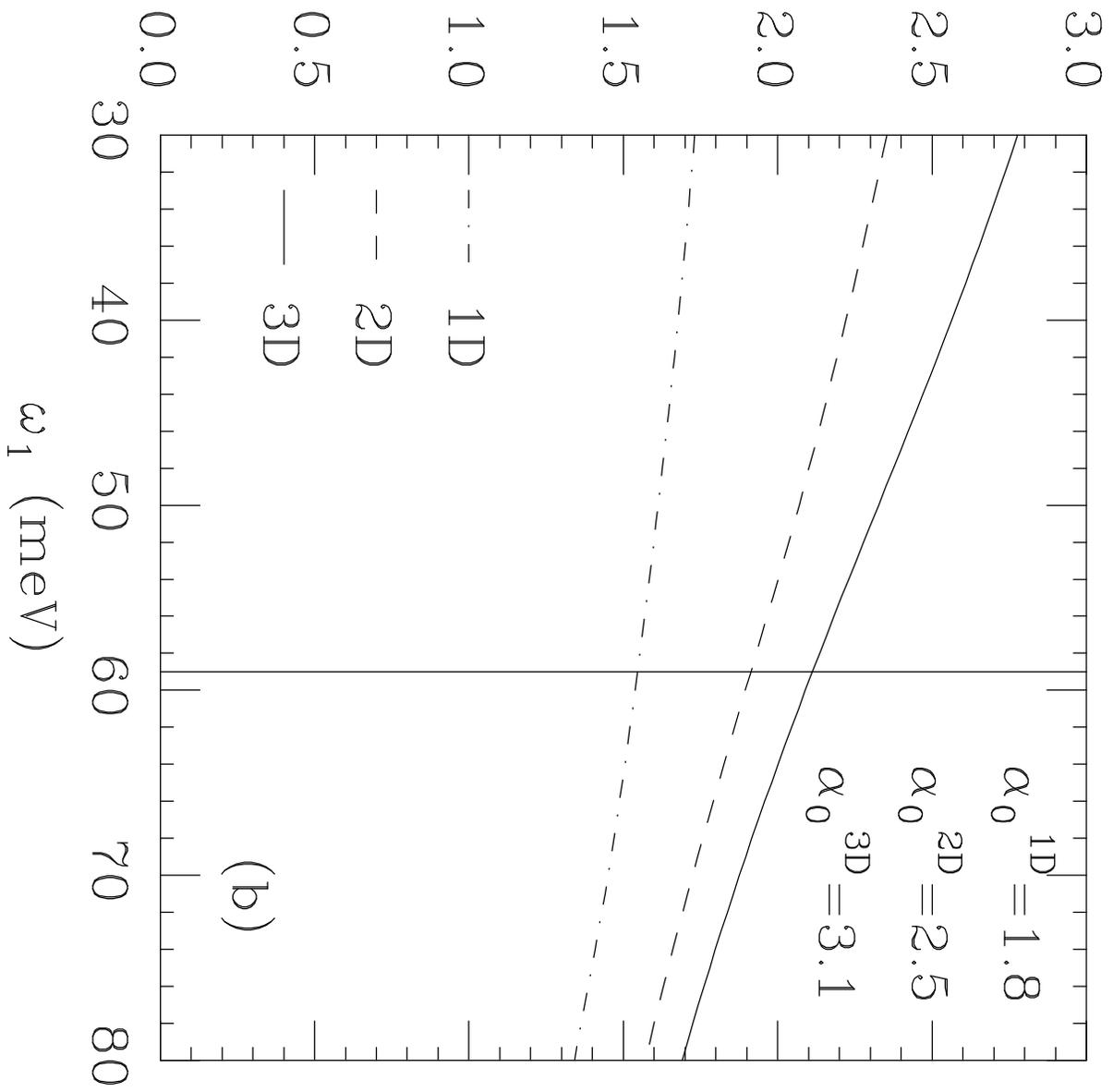
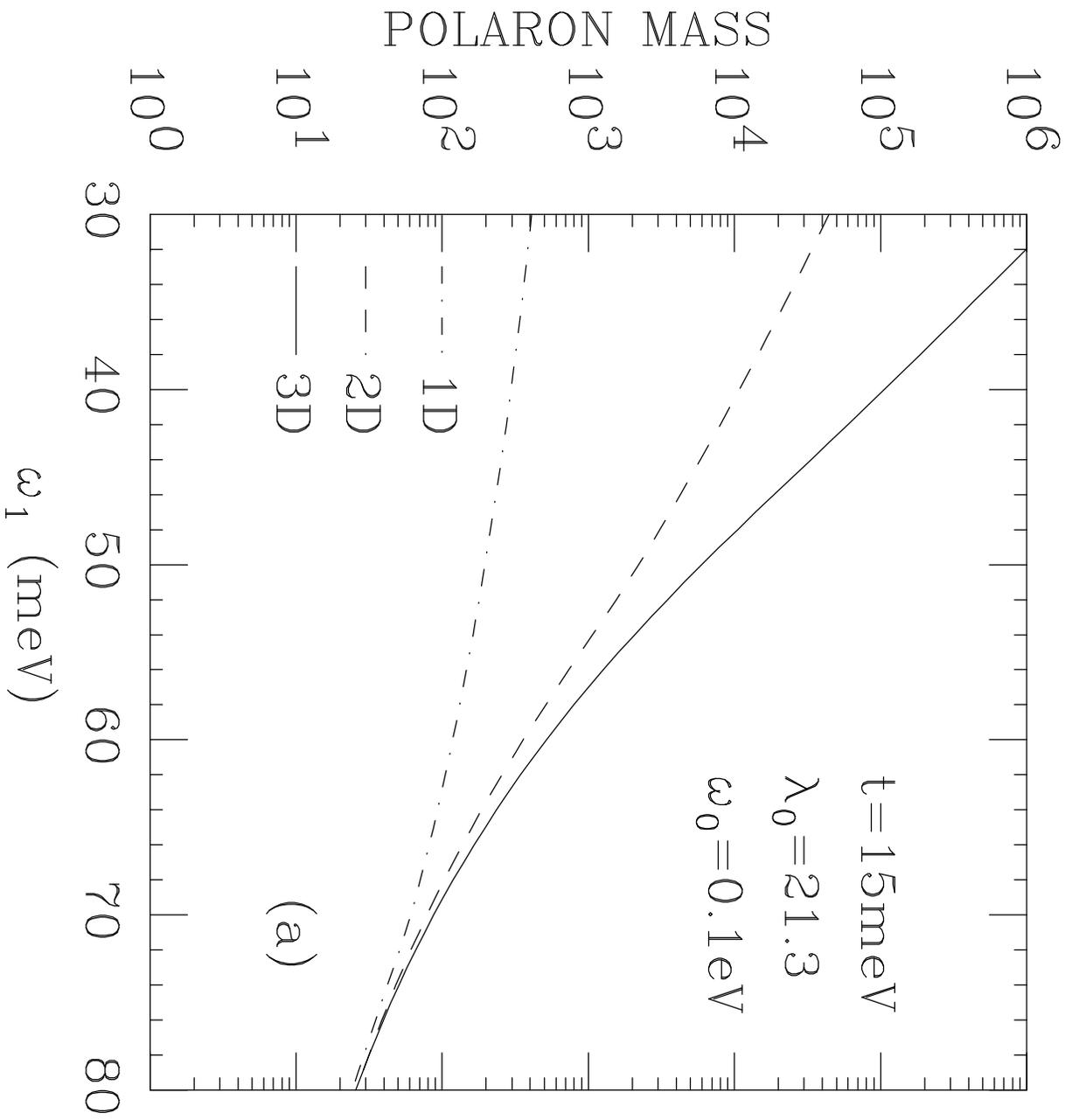
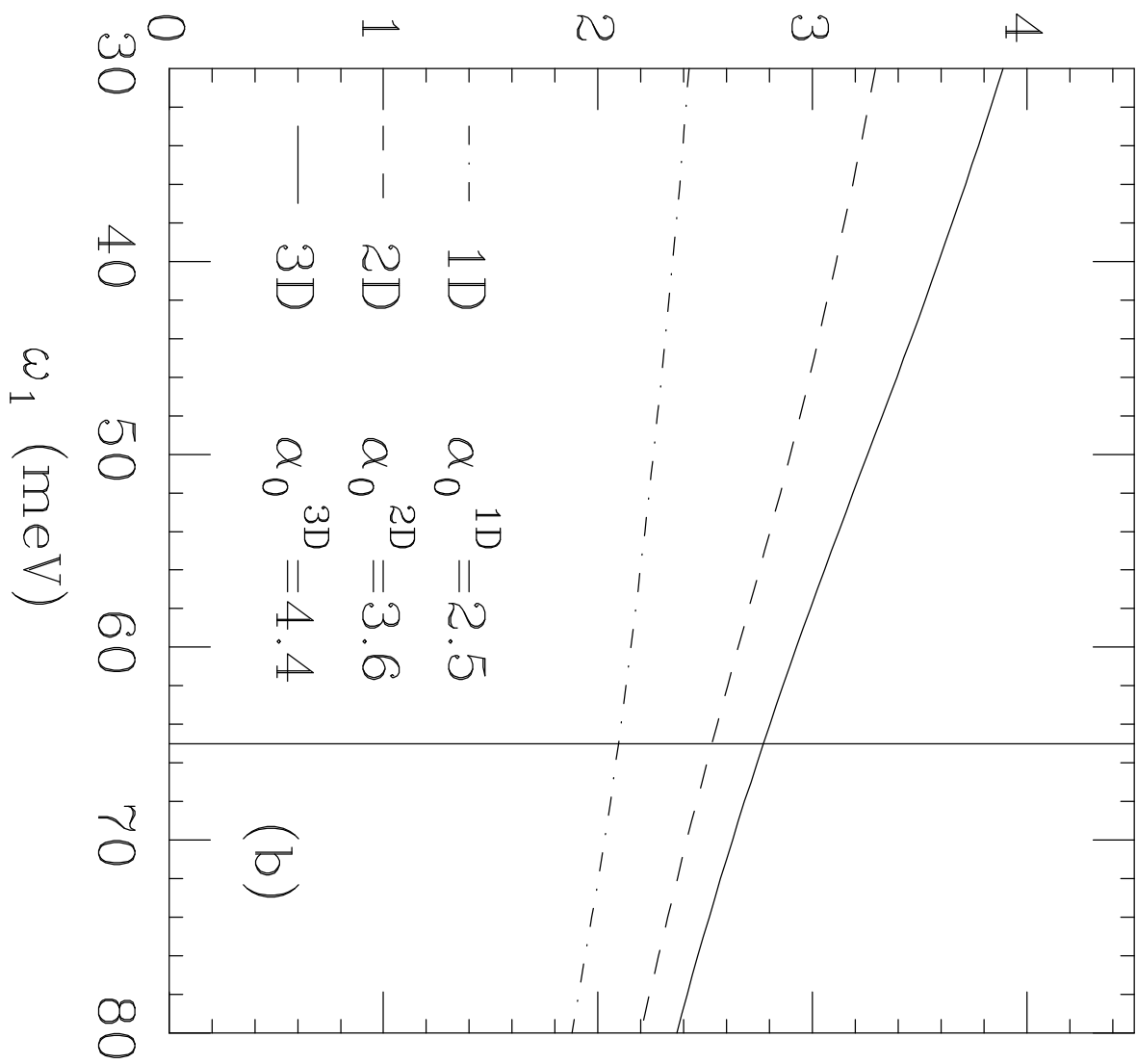
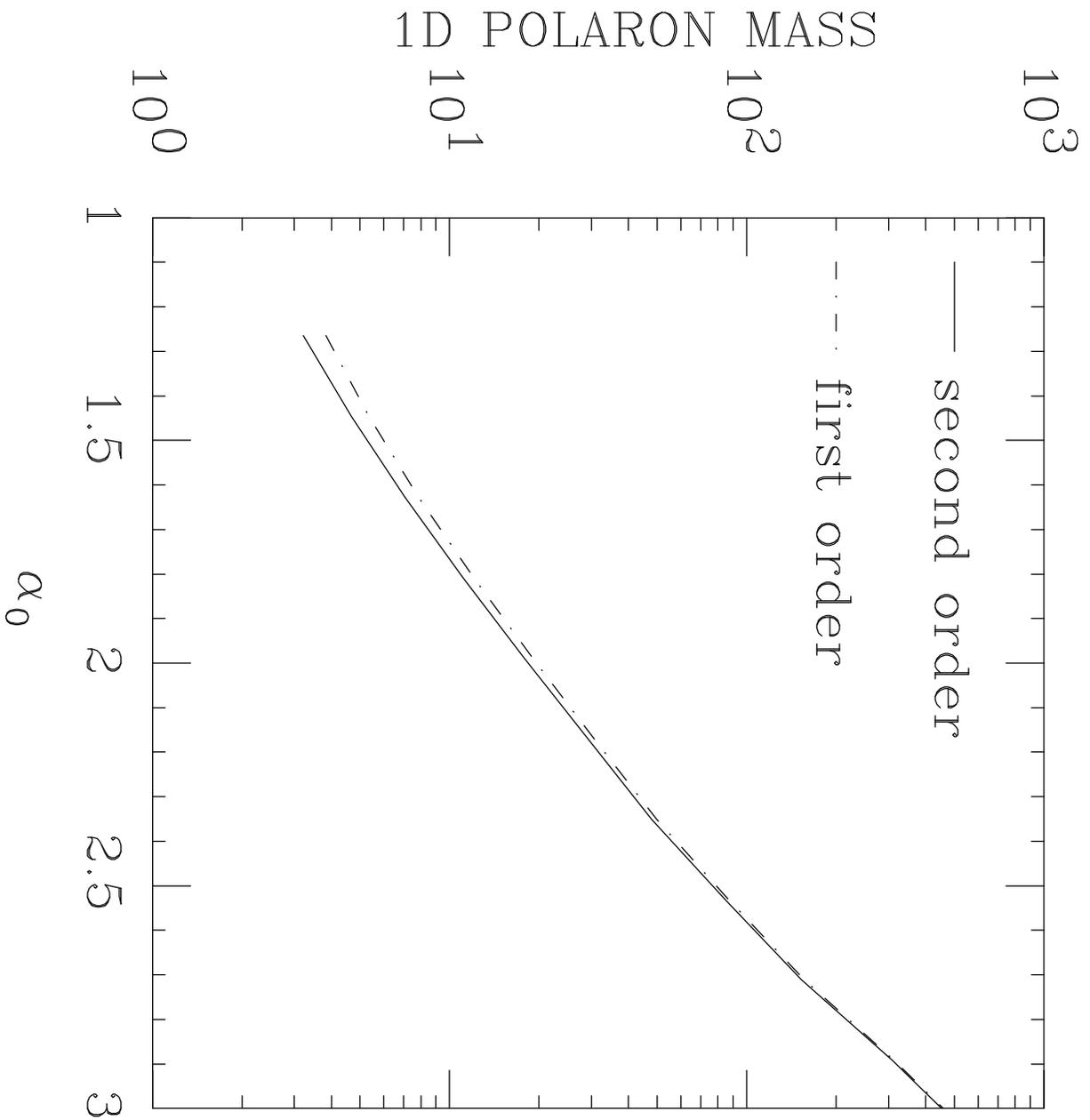
\end{document}